\documentclass[%
 reprint,
 amsmath,amssymb,
 prl,
floatfix,
]{revtex4-2}

\usepackage{graphicx}
\usepackage{dcolumn}
\usepackage{bm}
\usepackage[utf8]{inputenc}

\begin{document}

\title{A Chromomagnetic Mechanism for the Rotational Phase Transition of Gluonic Matter}

\author{Zhibin Li$^{1}$}
\email{lizhibin@zzu.edu.cn}
\author{Yidian Chen$^{2}$}
\email{chenyidian@hznu.edu.cn}
\author{Mei Huang$^{3}$}
\email{huangmei@ucas.ac.cn}

\affiliation{ $^{1}$ Institute for Astrophysics, School of Physics, Zhengzhou University, Zhengzhou 450001, China }
\affiliation{ $^{2}$ School of Physics, Hangzhou Normal University, Hangzhou, 311121, China}
\affiliation{ $^{3}$ School of Nuclear Science and Technology, University of Chinese Academy of Sciences, Beijing 100049, China
         }


\begin{abstract}
Rotation serves as a pivotal control parameter for QCD matter, yet effective models and lattice QCD yield conflicting predictions regarding its effect on the deconfinement transition. Using a rotation–magnetic correspondence within a holographic framework, we investigate the rotational response of pure gluonic matter. Calibrated against lattice QCD data at imaginary angular velocity, we find real rotation enhances chromomagnetic string tension and raises $T_c$, consistent with lattice QCD analytic-continuation predictions. The temperature dependence of chromomagnetic string tension dominates the system’s Barnett response: weak low-temperature tension induces the negative Barnett effect, and slightly above the transition, spin contributions prevail to generate an anomalous negative total moment of inertia. Since growing angular velocity further strengthens chromomagnetic string tension and suppresses spin-dominated inversion, this anomalous regime only survives at weak real rotation and vanishes at large angular velocity. At high temperature, fully restored strong string tension stabilizes conventional Barnett behavior. Stemming from the melting and thermal restoration of nonperturbative chromomagnetic flux tubes, our results establish the chromomagnetic-induced inertia inversion (CII) mechanism as the microscopic origin of this anomalous rotational response.
\end{abstract}

\maketitle

\textit{Introduction.}---
Mapping the phase structure of strongly interacting matter under extreme conditions is a central goal of nuclear physics. Beyond temperature and baryon density, rotation has emerged as a novel independent axis of the QCD phase diagram. Noncentral heavy-ion collisions generate substantial orbital angular momentum in the quark–gluon plasma, and experimentally observed hyperon polarization confirms partial conversion of this angular momentum into medium vorticity \cite{STAR:2017ckg,ALICE:2019aid,STAR:2022fan}. A key open question is whether rotation merely induces spin polarization or fundamentally modifies the QCD vacuum and deconfinement transition \cite{Jiang:2016woz,Kharzeev:2015znc}.

For pure gluonic matter, rotation-induced modifications of $T_c$ present a persistent theoretical contradiction. A broad class of effective models consistently predicts that rotation facilitates deconfinement and lowers $T_c$ \cite{Wang:2018sur,Fujimoto:2021xix,Chen:2020ath,Chernodub:2020qah,Zhao:2022uxc,Jiang:2023zzu,Jiang:2024zsw}. By contrast, lattice QCD is hindered by sign problems for real rotation but reliably accessible at imaginary angular velocity. Pure $SU(3)$ lattice simulations show that imaginary rotation reduces $T_c$, whose analytic continuation constrains real rotation corrections and implies an elevated $T_c$ under real rotation \cite{Braguta:2020biu,Braguta:2021jgn,Braguta:2022str,Yang:2023vsw}. Additional lattice and field-theory analyses further reveal near-transition anomalous rotational phenomena, including negative moment of inertia, rotational instability, and the negative Barnett effect \cite{Braguta:2023yjn,Braguta:2023tqz}. This discrepancy indicates that conventional effective models omit essential nonperturbative ingredients governing rotating gluonic vacuum dynamics.

This tension has stimulated extensive recent investigations. Improved strong-coupling analyses have advanced the understanding of imaginary-rotating gluodynamics \cite{Wang:2025mmv,Fukushima:2025hmh}, while complementary effective-model, holographic, and Yang–Mills studies have explored rotational effects in strongly coupled gauge theories \cite{Chen:2022smf,Chen:2022mhf,Sun:2024anu,Chen:2024jet,Chen:2026ced,Zhang:2026ctc}. Together with ongoing lattice efforts on rotating thermodynamics, finite-size corrections, and inhomogeneous vortical phases \cite{Yamamoto:2013zwa,Braguta:2023kwl,Braguta:2023qex,Braguta:2023iyx,Braguta:2024zpi}, these advances motivate a sign-problem-free nonperturbative framework enabling continuous analytic continuation from imaginary to real rotation.

Holographic QCD fulfills these requirements. Gauge–gravity duality provides nonperturbative, sign-problem-free access to strongly coupled gauge dynamics and permits straightforward analytic continuation in rotation parameters. Bottom-up Einstein–dilaton models accurately reproduce pure-glue and QCD thermodynamics \cite{GubserNellore:2008,Gubser:2008yx,GursoyKiritsis:2008,Gursoy:2008bu}. Extended Einstein–dilaton–Maxwell (EDM) constructions further capture nonperturbative magnetic-field effects on the QCD vacuum, including magnetized thermodynamics and phase structure \cite{Finazzo:2016mhm,Critelli:2016fvr,Dudal:2018ztm,Alho:2013hsa,Rougemont:2023gfz}, making the EDM framework suitable for rotating gluonic systems.

The rotation–magnetic correspondence establishes the necessary mapping. Originating from the Coriolis–Lorentz analogy and widely adopted in condensed-matter contexts \cite{Domenech:2010nf,Keranen:2009ss,Keranen:2009re,Dias:2013bwa,Xia:2019eje,Li:2019swh}, this effective dictionary maps rotational effects to magnetic-like field responses. Unlike conventional effective theories, this holographic approach reproduces lattice QCD rotational results and provides a unified description of the rotating gluonic vacuum. The key underlying mechanism is vorticity-induced modification of chromomagnetic fluctuations and nonperturbative flux-tube structures. Consequently, rotational behavior is governed not only by quasiparticle or Polyakov-loop dynamics but also by rotation-driven reshaping of the chromomagnetic vacuum condensate.

In this Letter, we resolve the conflict between effective models and lattice QCD by establishing a chromomagnetic-induced inertia inversion (CII) mechanism. Using a lattice-calibrated holographic framework, we show that the rotational response is governed by the competition between $J_{\rm orbital}$ and $J_{\rm spin}$ from chromomagnetic flux tubes. Near $T_c$, small chromomagnetic string tension drives $J_{\rm spin}$ antiparallel to $J_{\rm orbital}$, manifesting as a negative Barnett effect. In the range $T_c < T < T_s$ where $|J_{\rm spin}| > |J_{\rm orbital}|$, this leads to a negative total moment of inertia. For $T > T_s$, thermal recovery aligns $J_{\rm spin}$ parallel to $J_{\rm orbital}$, restoring the conventional positive Barnett effect. Real rotation enhances chromomagnetic stiffness, raising $T_c$ while suppressing the anomalous window, which vanishes at $v_c^2\simeq0.1$.

\begin{figure}[t]
    \centering
    \includegraphics[width=0.9\columnwidth]{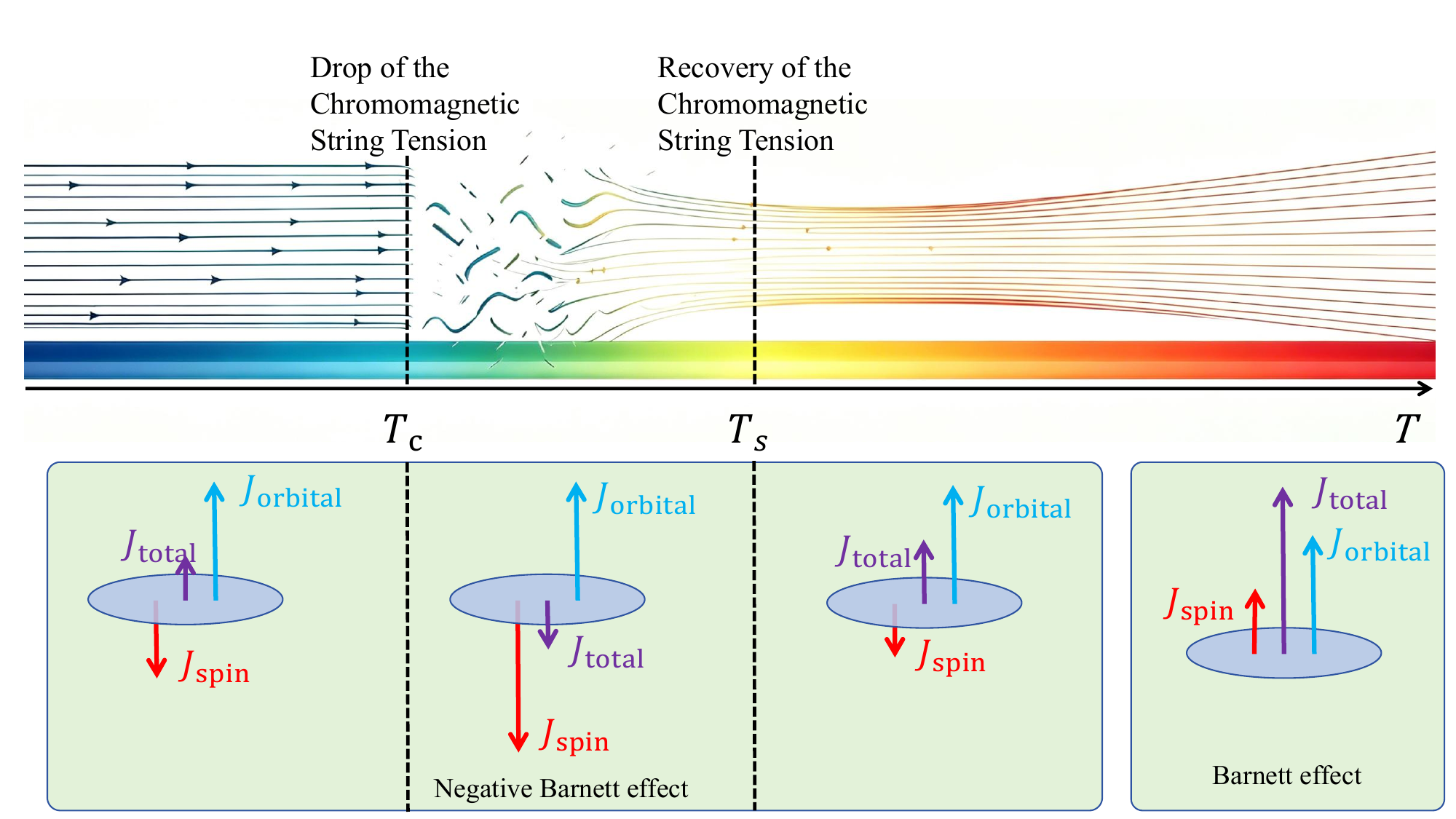}
    \caption{
    Schematic illustration of the chromomagnetic-induced inertia inversion (CII) mechanism. 
The total angular momentum decomposes as $J_{\rm total} = J_{\rm orbital} + J_{\rm spin}$. 
$J_{\rm orbital}$ (blue) is the kinematic baseline aligned with rotation $\Omega$, while $J_{\rm spin}$ (red) arises from chromomagnetic flux tube polarization. 
Near $T_c$, small chromomagnetic string tension drives $J_{\rm spin}$ antiparallel to $J_{\rm orbital}$ (negative Barnett effect). 
In the range $T_c < T < T_s$, where $|J_{\rm spin}| > |J_{\rm orbital}|$, this yields a negative total moment of inertia ($J_{\rm total} < 0$). 
For $T > T_s$, thermal recovery aligns $J_{\rm spin}$ parallel to $J_{\rm orbital}$, restoring the conventional positive Barnett effect ($J_{\rm total} > 0$).
    }
    \label{fig:schematic}
\end{figure}

\textit{Holographic setup and rotational dictionary.}--- 
To describe the rotational dynamics of pure-gluon matter, we adopt the rotation--magnetic correspondence as an effective long-wavelength dictionary. Inspired by the Coriolis--Lorentz-force analogy and effective gauge-field responses in rotating condensed-matter systems \cite{Domenech:2010nf,Keranen:2009ss,Keranen:2009re,Dias:2013bwa,Xia:2019eje,Li:2019swh}, this approach maps rotational kinematics onto magnetic dynamics. Although it is not an exact microscopic equivalence between relativistic rotation and electromagnetism, it serves as a phenomenological bridge. Concretely, for a system of finite radius $R$ with boundary velocity $v=R\Omega<1$ at fixed temperature, we map a rotation about the $z$-axis to a homogeneous bulk magnetic field $B=2m(T)\Omega$. The five-dimensional holographic action is
\begin{equation}
\begin{aligned}
S=\frac{1}{2\kappa_5^2}
\int d^5x\sqrt{-\mathcal{G}}
&\left[
\mathcal{R}-\frac{1}{2}\partial_\mu\phi\partial^\mu\phi
-V(\phi)
\right.
\\
&\left.
-\frac{Z(\phi)}{4}F_{\mu\nu}F^{\mu\nu}
\right],
\end{aligned}
\end{equation}
where $\kappa_5^2$ is the five-dimensional gravitational constant, $\mathcal{G}$ is the determinant of the metric tensor, $\mathcal{R}$ is the Ricci scalar (distinct from the system radius $R$ introduced above), and $F_{\mu\nu}$ is the field strength tensor of the bulk gauge field $A_\mu$ (with $\mu,\nu$ running over the five bulk coordinates). Here $\phi$ is the dilaton field with potential $V(\phi)$, while $Z(\phi)$ controls the coupling of the magnetic field to the nonperturbative gluonic sector.

Rather than fine-tuning parameters for exact quantitative agreement with lattice QCD, we prioritize a minimalist and physically transparent framework. By keeping the model inputs minimal, we robustly capture the essential qualitative features of the system, ensuring that the underlying physical mechanisms are not obscured by unnecessary structural complexity. Specifically, we adopt $V(\phi)=-12\cosh(\gamma \phi)+\left(6\gamma^2-\frac{3}{2}\right)\phi^2$ and $Z(\phi)=\mathrm{sech}(\zeta\phi^3)$, with nonrotating parameters fixed at $\gamma=0.735$, $\kappa_5^2=9.76\pi$, and $\zeta=0.1275$ \cite{He:2022amv}. Related EMD and improved-holographic-QCD studies establish the thermodynamic, transport, and magnetic-field benchmarks against which such effective backgrounds are normally assessed \cite{Gursoy:2008bu,Gubser:2008yx,Alho:2013hsa,Finazzo:2016mhm,Rougemont:2023gfz,Dudal:2018ztm,Critelli:2016fvr}. For the effective scale $m(T)$, we adopt the linear ansatz $m(T)=\lambda T$, physically motivated by the characteristic linear temperature scaling of the gluon Debye mass ($m_\text{D} \sim g_s T$) and magnetic mass ($m_\text{mag} \sim g_s^2 T$) \cite{Arnold:1995bh,Xu:2008av,Silva:2013maa}, where $g_s$ is the strong coupling constant. The dimensionless mapping parameter $\lambda$ is calibrated against the low-velocity response at imaginary angular velocity.

For a magnetic field along the rotation axis, we use the anisotropic ansatz
\begin{equation}
ds^2
=
-e^{-\eta(r)}f(r)dt^2
+\frac{dr^2}{f(r)}
+r^2(dx^2+dy^2)
+r^2g(r)dz^2 ,
\end{equation}
where $t, r, x, y, z$ are the coordinates, $f(r)$ and $\eta(r)$ are metric functions, and $g(r)$ encodes the breaking of spatial rotational symmetry between the transverse plane $(x,y)$ and the longitudinal direction $z$. The bulk gauge field is given by $A_\mu dx^\mu=-m \Omega( y\,dx - x\,dy)$. This magnetic-field-induced anisotropic geometry is the Einstein--Maxwell--dilaton analogue of the magnetic-brane construction \cite{DHokerKraus:2009}.

The temperature and entropy density are determined by the horizon data,
\begin{equation}
T=\frac{f'(r_h)}{4\pi},
\qquad
s=\frac{2\pi}{\kappa_5^2}r_h^3 ,
\end{equation}
where $r_h$ is the horizon radius (the largest root of $f(r_h)=0$) and $f'(r_h)$ is its derivative with respect to $r$. The deconfinement temperature $T_c$ is obtained from the first-order transition between competing black-hole branches, equivalently from the discontinuity in entropy and the pressure crossing.

The boundary stress tensor yields the energy density and anisotropic pressures. Under the rotation--magnetic correspondence, the pressure anisotropy satisfies $P_\parallel-P_\perp=\Omega J_{\rm total}$ with the angular momentum density $J_{\rm total}=\partial P_{\parallel}/\partial \Omega$, the dual counterpart of the thermodynamic relation $P_\parallel-P_\perp=MB$ for magnetized media \cite{Bali:2014kia,Karmakar:2019tdp,Chaudhuri:2022oru}. Here $P_\parallel$ and $P_\perp$ are the longitudinal and transverse pressures, $J$ is the angular momentum density, $M$ is the magnetization, and $B$ is the magnetic induction. This relation allows us to extract the rotational susceptibility directly from the holographic anisotropic pressure.

Near the AdS boundary, the anisotropy function admits the asymptotic expansion
\begin{equation}
 g(r)
=
g_0+\frac{g_v}{r^4}
-\frac{m^2 \Omega^2 g_0Z(0)\ln r}{r^4}
+\cdots ,
\end{equation}
where $g_0$, the boundary value of the anisotropy function, is set to $g_0=1$, and $g_v$ is the normalizable coefficient encoding the expectation value of the anisotropic stress. The pressure anisotropy then takes the form
\begin{equation}
P_\parallel-P_\perp
=
\frac{1}{2\kappa_5^2}
\left[
m^2 \Omega^2 Z(0)+\frac{4g_v}{g_0}
\right] =\Omega J_{\rm total}.
\end{equation}
The $\Omega J_{\rm total}$ can be naturally decomposed into two components. Accordingly, the total moment of inertia $I_{\rm total}=J_{\rm total}/\Omega$ can be similarly separated as
\begin{equation}
 I_{\rm total}=I_{\rm orbital}+I_{\rm spin},
\end{equation}
with $I_{\rm orbital}=\frac{1}{2\kappa_5^2}m^2Z(0)$ and $I_{\rm spin}=\frac{1}{2\kappa_5^2}\frac{4g_v}{g_0\Omega^2}$. The first term represents the positive rigid-rotation contribution, while the second captures the anisotropic, nonperturbative response encoded by $g_v$. Within this effective dictionary, we refer to the latter as a spin contribution, though the unambiguous observable remains the total rotational response.

This decomposition is naturally realized within the holographic dictionary: the logarithmic term in the expansion of $g(r)$ is fixed entirely by the external source (kinematic response), while the normalizable coefficient $g_v$ strictly encodes the vacuum expectation value of the dual operator, capturing the dynamical reorganization of internal degrees of freedom. The correspondence is restricted to the low-velocity, specified-radius regime stated above. Agreement with data not used in fixing $\lambda$, and sensitivity to $R$ and to the magnetic coupling $Z(\phi)$, provide the appropriate tests of this effective description.

\begin{figure}[t]
    \centering
    \includegraphics[width=0.9\columnwidth]{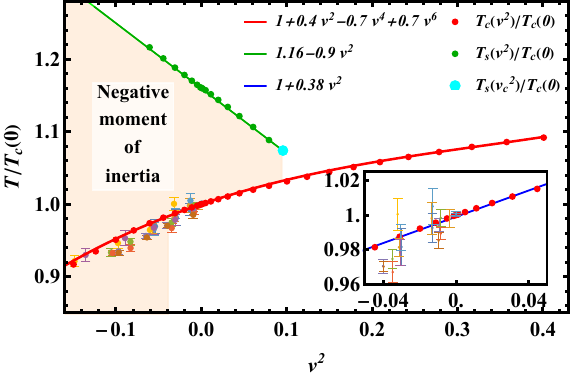}
    \caption{
    Deconfinement temperature as a function of squared rotational velocity $v^2$. The data points with error bars correspond to lattice QCD results \cite{Braguta:2021jgn}. Negative $v^2$ corresponds to imaginary angular velocity, where direct lattice data are available, while positive $v^2$ gives the analytic continuation to real rotation. The holographic result reproduces the lattice suppression of $T_c$ at imaginary rotation and predicts an enhancement under real rotation. The low-velocity behavior follows $T_c(v^2)/T_c(0)\simeq 1+0.38v^2$, whereas larger real velocities require higher-order corrections. The green curve denotes the supervortical temperature $T_s$, which vanishes at $v_c^2\simeq0.1$.}
    \label{fig:Tc}
\end{figure}

\textit{Rotational deconfinement and analytic continuation.}--- 
We first consider imaginary angular velocity, $v^2<0$, where direct lattice simulations are available. Adopting the finite-system convention $R=1\,\mathrm{fm}$, we set $\lambda=11.8$ by fitting the low-velocity rotational dependence of $T_c$. Through this calibration, we confirm the lattice results regarding the suppression of $T_c$ under imaginary rotation (Fig.~\ref{fig:Tc}), validating our framework for analytic continuation to real angular velocity.

Within the analytic domain of the present model, the imaginary-rotation trend continues to an enhancement of $T_c$ for real rotation. For small $v^2$, the transition temperature follows $\frac{T_c(v^2)}{T_c(0)}\simeq1+0.38v^2$, consistent with the low-velocity scaling inferred from lattice QCD. As the real rotational velocity increases, the dependence becomes nonlinear (Fig.~\ref{fig:Tc}), well approximated by $\frac{T_c(v^2)}{T_c(0)}=1+0.4v^2-0.7v^4+0.7v^6$. The quadratic approximation is reliable only near the nonrotating limit; higher-order terms constitute a model-dependent extrapolation for the real-rotation domain, providing benchmarks for future first-principles lattice simulations.

The pressure anisotropy also reveals a second characteristic temperature $T_s$ (green curve in Fig.~\ref{fig:Tc}), marking the point where the rotational response changes sign, $I_{\rm total}(T_s,\Omega)=0$. At small real rotation, $T_s\simeq 1.16 T_c$. As real rotation increases, $T_s$ decreases approximately as $\frac{T_s(v^2)}{T_c(0)}\simeq1.16-0.9 v^2$. The anomalous regime disappears at a critical velocity $v_c^2\simeq 0.1$, above which the pressure response becomes conventional over the entire temperature range.

\textit{Anomalous rotational response near deconfinement.}--- 
Within our holographic rotation–magnetic dictionary, the total angular-momentum response decomposes into orbital and spin contributions, $J_{\rm orbital}=I_{\rm orbital}\Omega$ and $J_{\rm spin}=I_{\rm spin}\Omega$. The orbital term is always positive, characterizing rigid rotational motion of the gluonic medium. In contrast, the spin contribution strongly depends on temperature and rotation strength. This dependence yields a well-defined two-regime Barnett behavior: the negative Barnett effect prevails at low temperatures below approximately $1.5\,T_c$, while the conventional Barnett behavior dominates at high temperatures. Our analytic continuation from the near-zero rotation limit confirms that this temperature-dependent anomalous inertia persists for physical real rotation.

Crucially, the total moment of inertia increases monotonically with $v^2$, which reshapes the temperature range of the negative $I_{\rm total}$ regime. For rotation with $v^2\lesssim-0.04$, the negative spin contribution dominates throughout the entire low-temperature region, yielding $I_{\rm total}<0$ for all $T\lesssim T_s$. As $v^2$ increases, the growing total inertia suppresses the low-temperature anomalous behavior, restricting negative $I_{\rm total}$ exclusively to a finite near-transition window $T_c\lesssim T\lesssim T_s$. Below $T_c$, the system exhibits positive total moment of inertia. Immediately above $T_c$, the spin contribution dominates and generating anomalous negative inertia. For $T>T_s$, spin susceptibility recovers to positive values, terminating the negative Barnett regime.

\begin{figure}[t]
    \centering
    \includegraphics[width=0.9\columnwidth]{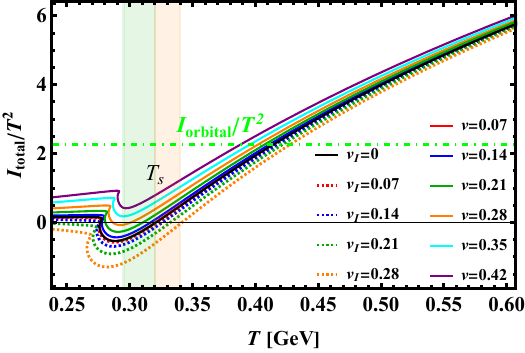}
    \caption{Total moment of inertia $I_{\rm total}/T^2$ as a function of temperature under varying rotation strengths. The constant positive orbital contribution $I_{\rm orbital}/T^2$ is shown for reference. The system exhibits negative Barnett behavior at low temperatures ($T\lesssim 1.5\,T_c$) and conventional Barnett behavior at high temperatures. Due to the monotonic increase of total inertia with $v^2$, negative $I_{\rm total}$ persists over the full low-temperature range for $v^2\lesssim-0.04$, while it is confined to the finite window $T_c\lesssim T\lesssim T_s$ for $v^2>-0.04$. The anomalous window fully vanishes at $v^2\simeq0.1$.}
    \label{fig:inertia}
\end{figure}

Further increasing $v^2$ continuously suppresses the anomalous inertial window. A larger $v^2$ raises $T_c$ and lowers the upper threshold $T_s$, steadily narrowing the finite temperature interval that supports negative $I_{\rm total}$ (Fig.~\ref{fig:inertia}). While the orbital contribution remains uniformly positive, the spin-induced negative inertia only dominates within the shrinking near-transition regime. Once $v^2$ exceeds the critical value $v_c^2\simeq0.1$, the monotonic enhancement of total inertia fully eliminates the negative spin contribution, closing the anomalous window completely and restoring conventional Barnett behavior at all temperatures.

\textit{Chromomagnetic string tension and microscopic origin.}--- 
The negative Barnett effect is directly rooted in the nonperturbative chromomagnetic structure of the gluonic medium. To explicitly establish this physical connection, we compute the spatial Wilson loop and extract the corresponding chromomagnetic string tension within our holographic framework. The transverse and longitudinal chromomagnetic string tensions read $\sigma_{\perp}^B=\frac{g_p}{\pi}r_h^2 \mathrm{e}^{\sqrt{\frac{2}{3}}\phi(r_h)}$ and $\sigma_\parallel^B=\sigma_\perp^B\sqrt{g(r_h)}$, where $g_p$ is a normalization constant fixed by matching the nonrotating spatial string tension to pure-gauge lattice data \cite{KarschLaermannLuetgemeier:1995}, and $\phi(r_h)$ denotes the dilaton field evaluated at the black hole horizon.

\begin{figure}[t]
    \centering
    \includegraphics[width=0.9\columnwidth]{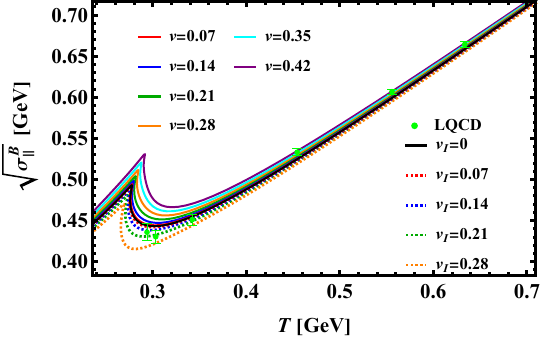}
    \caption{Longitudinal chromomagnetic string tension $\sqrt{\sigma_\parallel^B}$ at different angular velocities. The holographic result agrees with pure-gauge lattice data in the nonrotating limit \cite{Boyd:1996bx}. Its temperature and rotational dependence closely track the spin contribution to the moment of inertia: weak chromomagnetic string tension yields the negative Barnett effect, while restored large string tension produces conventional Barnett behavior, both driven by the melting and thermal restoration of nonperturbative chromomagnetic flux tubes.}
    \label{fig:string}
\end{figure}

The temperature dependence of $\sqrt{\sigma_\parallel^B}$ reproduces the characteristic spatial string tension of pure $SU(3)$ gluodynamics in the nonrotating limit, in agreement with lattice results \cite{Boyd:1996bx}. Importantly, this observation supplies direct evidence supporting the chromomagnetic-induced inertia inversion (CII) mechanism. The string tension displays pronounced rotational dependence and closely follows the spin contribution to the total moment of inertia (Fig.~\ref{fig:string}). It falls rapidly near $T_c$ concurrent with the suppression of spin angular momentum, and gradually rises at higher temperatures as the nonperturbative chromomagnetic sector undergoes thermal restoration. This tight coevolution verifies that the CII mechanism governs the anomalous rotational response of the gluonic medium: melting of chromomagnetic flux tubes produces weak string tension and thus the negative Barnett effect, while thermal restoration regenerates strong string tension that recovers the conventional positive Barnett behavior, with both regimes fully dictating the magnitude and sign of the spin contribution.

This microscopic mechanism also resolves the opposite effects of real and imaginary rotation on $T_c$: imaginary rotation suppresses the chromomagnetic sector and lowers the deconfinement temperature, whereas real rotation enhances chromomagnetic stiffness and elevates $T_c$ upon analytic continuation. The longstanding discrepancy between lattice QCD simulations and conventional effective models is resolved once the chromomagnetic sector is treated as a dynamical, nonperturbative component of the rotating gluonic vacuum.

\textit{Conclusion.}---
Using lattice-calibrated rotation–magnetic correspondence, we investigate the rotational response of pure gluonic matter and identify the chromomagnetic-induced inertia inversion (CII) mechanism for anomalous vortical behavior. We demonstrate that the nonperturbative chromomagnetic sector governs the temperature and rotation dependence of gluonic inertial responses. Constrained by low-velocity imaginary-rotation lattice data, our physical description reproduces key lattice features and enables reliable analytic continuation to real rotation. The results show that real rotation raises $T_c$ with large-velocity nonlinear corrections, and the negative Barnett effect occurs only under weak real rotation. Increasing rotation strengthens chromomagnetic string tension, suppresses spin-driven inertial inversion, and completely removes the anomaly at $v_c^2\simeq0.1$.

The synchronized evolution of chromomagnetic string tension and rotational response confirms that the CII mechanism arises from temperature-dependent melting and restoration of chromomagnetic flux tubes. Near $T_c$, melted flux tubes produce weak string tension and negative Barnett behavior, while thermally restored strong tension recovers conventional positive Barnett responses. This picture unifies rotation-modulated deconfinement and anomalous rotational dynamics in pure gluodynamics, offering a sign-problem-free nonperturbative explanation for the distinct behaviors of imaginary and real rotation. This work paves the way for further numerical validation and can be extended to dynamical quark systems to explore CII-induced vortical phenomena near the QCD crossover.

\begin{acknowledgments}
We thank Yu Tian and Chao Wu for useful discussions. This work was supported in part by the Natural Science Foundation of Henan Province of China under Grant No. 262300421877, the National Natural Science Foundation of China (NSFC) Grant No: 12305136, 12235016, 12221005 and the start-up funding of Hangzhou Normal University under Grant No. 4245C50223204075.
\end{acknowledgments}

\bibliographystyle{apsrev4-2}

%

\clearpage
\onecolumngrid

\begin{center}
{\large Supplemental Material}
\end{center}

\setcounter{equation}{0}
\renewcommand{\theequation}{S\arabic{equation}}

\textit{Holographic renormalization and thermodynamic dictionary}---
To implement the rotation-magnetic correspondence in the holographic framework, we introduce an effective bulk magnetic field into the 5D Einstein-dilaton-Maxwell action:
\begin{eqnarray}
    S=\frac{1}{2\kappa_5^2}\int d^5x \sqrt{-g}\left[\mathcal{R}-\frac{1}{2}\nabla_{\mu}\phi\nabla^{\mu}\phi-V(\phi)-\frac{Z(\phi)}{4}F^2\right] .
\end{eqnarray}
Here $\mathcal{R}$ is the Ricci scalar. The field $\phi$ is the dilaton field, which is dual to the gluon condensate operator. $F$ is the bulk magnetic field which corresponds to the rotation. $Z(\phi)$ is the coupling of the dilaton to the magnetic fields, controlling the running of the gauge coupling. Following the common convention, we impose the rotation along the $z$-direction. Then the metric and gauge fields can be set as 
\begin{equation}
\begin{aligned}
ds^2=&-e^{-\eta(r)}f(r)dt^2+\frac{1}{f(r)}dr^2+r^2(dx^2+dy^2)+r^2 g(r)dz^2,\\ 
  &  A_{\mu}dx^{\mu}=-\frac{B}{2}ydx+\frac{B}{2}xdy,~~~F=d A,
\end{aligned}
\end{equation}
where the anisotropic factor $g(r)$ captures the spatial symmetry breaking induced by rotation, the effective magnetic field $B=2m(T) \Omega$, and $e^{-\eta(r)}$ is the warp factor. Varying the action with respect to the bulk fields yields the coupled equations of motion:
\begin{equation}
\begin{aligned}
   \phi''+\left( \frac{f'}{f}+\frac{3}{r}+\frac{g'}{2g}-\frac{\eta'}{2} \right)\phi'-\frac{V'}{f}-\frac{B^2 Z'}{2r^4 f}&=0,\\ 
    g''+\left( \frac{2}{r}+\frac{\eta'}{2} \right)g'-\frac{g'^2}{2g}+\left( \frac{3\eta'}{r}+\phi'^2 \right)g&=0,\\ 
    \frac{B^2 }{r^3}Z g-f g'-r f' g'+3 f g \eta'+r f g' \eta'+r f g \phi'^2&=0,\\ 
    \frac{4}{r^2}+\frac{2V}{3f}+\frac{B^2 Z}{3 r^4 f}+\frac{2f'}{rf}+\frac{4g'}{3rg} +\frac{f'g'}{3fg}-\frac{2\eta'}{r}-\frac{g'\eta'}{3g}-\frac{\phi'^2}{3}&=0.
\end{aligned}
\end{equation}

To render the on-shell action finite and extract well-defined boundary observables, we employ the standard holographic renormalization procedure by adding the boundary counterterm action at the UV cutoff $r \to \infty$:
\begin{equation}
\begin{aligned}
   S_\partial =\frac{1}{2\kappa_5^2}\int_{r\rightarrow\infty} d^4x \sqrt{-h} & \left\{2\mathcal{K}-12-\frac{1}{2}\phi^2 +\frac{1}{6}V'''(0) \phi^3-b \phi^4 \right. \\
    &\left. +\left[\frac{1}{24}\left(2-3V'''(0)^2+V^{(4)}(0) \right)\phi^4+\frac{1}{4}Z(0)F^2\right]\ln(r)\right\},
\end{aligned}
\end{equation}
with $h$ the determinant of the induced metric at boundary and $\mathcal{K}$ the extrinsic curvature defined by the outward pointing normal vector to the boundary. The terms in the braces are designed to cancel the power-law and logarithmic divergences. $b$ is an integration constant fixing the free energy at vanishing temperature and chemical potential, which removes the scheme dependence of the renormalization. 

At the IR boundary, we impose regularity conditions near the black horizon:
\begin{eqnarray}
    X|_{r\rightarrow r_h}=X_0+X_1(r-r_h)+X_2(r-r_h)^2+\dots ,
\end{eqnarray}
for $X=f,~\eta,~\phi,~g$. And $f_0=0$ corresponds to the coordinate singularity at the black hole horizon, defining the Hawking temperature. Choosing the gauge $\eta(r_h)=0$ to ensure the proper periodicity of the Euclidean time, the Hawking temperature and black hole entropy can be calculated as
\begin{eqnarray}
   T&=&\frac{1}{4\pi}\sqrt{e^{-\eta(r_h)}f'(r_h)\left[f'(r_h)-f(r_h)\eta'(r_h)\right]}=\frac{1}{4\pi} f'(r_h),\\
    S_{BH}&=&\frac{2\pi}{\kappa_5^2}\int_{r=r_h}\sqrt{\tilde{h}}dxdydz=\frac{2\pi}{\kappa_5^2}\tilde{V}r_h^3,
\end{eqnarray}
where $\tilde{h}$ is the determinant of the induced metric on the horizon as $\tilde{h}_{ab}dx^adx^b=r^2 d\vec{x}^2$. The $\tilde{V}$ is the volume of the total space. And the entropy density is given by
\begin{equation}
    s=\frac{2\pi}{\kappa_5^2}r_h^3.
\end{equation}

Solving the equations of motion asymptotically near the AdS boundary, the UV expansions of the bulk fields take the form
\begin{equation}
\begin{aligned}
f|_{r\rightarrow \infty}&=r^2+\frac{\phi_s^2}{6}-\frac{2\phi_s^3V'''(0)}{9r}+\frac{f_v}{r^2}-\left[12B^2Z(0)+\left(2-3V'''(0)^2+V^{(4)}(0)\right)\phi_s^4\right]\frac{\ln(r)}{24r^2}+\mathcal{O}(r^{-3}),\\ 
    \eta|_{r\rightarrow \infty}&=\eta_0+\frac{\phi_s^2}{6r^2}-\frac{2\phi_s^3 V'''(0)}{9r^3}+\mathcal{O}(r^{-4}),\\ 
    \phi|_{r\rightarrow \infty}&=\frac{\phi_s}{r}-\frac{\phi_s^2 V'''(0)}{2r^2}+\frac{\phi_v}{r^3}-\left(2-3 V'''(0)^2+V^{(4)}(0)\right)\frac{\phi_s^3\ln(r)}{12 r^3}+\mathcal{O}(r^{-4}),\\ 
    g|_{r\rightarrow \infty}&=g_0+\frac{g_v}{r^4}-\frac{B^2 g_0Z(0) \ln(r)}{4r^4}+\mathcal{O}(r^{-5}).\\ 
\end{aligned}
\end{equation}
According to the AdS/CFT dictionary, the subscripts $s$ and $v$ denote the source (non-normalizable mode) and the vacuum expectation value (normalizable mode) of the dual boundary operators, respectively. 

Finally the thermal quantities can be read from the expectation value of the renormalized energy-momentum tensor on the boundary as
\begin{equation}
\begin{aligned}\label{eq11}
    \epsilon&=T_{tt}=\frac{1}{2\kappa_5^2}\left[-3f_v+\frac{4g_v}{g_0}+\left(\frac{1}{48}+b\right)\phi_s^4+\phi_s\phi_v+\frac{1}{4}B^2 Z(0) +\frac{3}{8}V'''(0)^2\phi_s^4\right],\\ 
    P_{\perp}&=T_{xx}=T_{yy}=\frac{1}{2\kappa_5^2}\left[-f_v+\phi_s\phi_v-\frac{1}{6}B^2 Z(0)-\frac{1}{144}\left( 144b-9+12V'''(0)^2-2V^{(4)}(0)\right)\phi_s^4\right],\\ 
    P_{\parallel}&=T_{zz}=\frac{1}{2\kappa_5^2}\left[-f_v+\frac{4g_v}{g_0}+\phi_s\phi_v+\frac{1}{12}B^2 Z(0)-\frac{1}{144}\left( 144b-9+12V'''(0)^2-2V^{(4)}(0)\right)\phi_s^4\right],\\ 
    \theta&=\epsilon-2P_{\perp}-P_{\parallel}=\frac{1}{2\kappa_5^2}\left[  -2\phi_s\phi_v+\frac{1}{2} B^2 Z(0)-\frac{1}{24}\left( 4-96b-15V'''(0)^2 +V^{(4)}(0) \right)\phi_s^4 \right],
\end{aligned}
\end{equation}
where $\theta$ denotes the trace anomaly of the boundary field theory. Under the rotation-magnetic mapping, the thermodynamic variables satisfy the relation
\begin{eqnarray}
    \epsilon=\epsilon_{\rm total}-\Omega J_{\rm total},~~ P_{\parallel}-P_{\perp}=\Omega J_{\rm total},
\end{eqnarray}
where $J_{\rm total}=\text{d}P_{\parallel}/\text{d}\Omega$ is the angular momentum density, and $\Omega$ represents the angular velocity.

Evaluating the pressure anisotropy from Eq.~(\ref{eq11}), we have 
\begin{eqnarray}
    P_{\parallel}-P_{\perp}=\frac{1}{2\kappa_5^2}\left( \frac{1}{4}B^2 Z(0)+ \frac{4g_v}{g_0}\right)=\Omega J_{\rm total}.
\end{eqnarray}
The first term 
\begin{eqnarray}
    \Omega J_{\rm orbital}=\frac{1}{2\kappa_5^2}\frac{1}{4}B^2 Z(0)=\frac{1}{2\kappa_5^2}m^2 Z(0) \Omega^2=I_{\rm orbital}\Omega^2,
\end{eqnarray}
is proportional to $\Omega^2$ which corresponds to the contribution of orbital angular momentum as $J_{\rm orbital}=I_{\rm orbital}\Omega$ with $I_{\rm orbital}=\frac{1}{2\kappa_5^2}m^2 Z(0)$. Here, the factor $Z(0)$ arises from the dilaton-gauge coupling evaluated at the UV boundary. And the second term
\begin{eqnarray}
    \Omega J_{\rm spin}=\frac{1}{2\kappa_5^2}\frac{4g_v}{g_0},
\end{eqnarray}
corresponds to the contribution of spin angular momentum from spin alignment, which is governed by the anisotropic metric deformation $g_v$. Then the total moment of inertia is
\begin{eqnarray}
    I_{\rm total}=I_{\rm orbital}+I_{\rm spin}=\frac{1}{2\kappa_5^2}m^2 Z(0)+\frac{1}{2\kappa_5^2}\frac{4g_v}{g_0 \Omega^2}.
\end{eqnarray}

\begin{figure}[t]
    \centering
    \includegraphics[width=.43\textwidth]{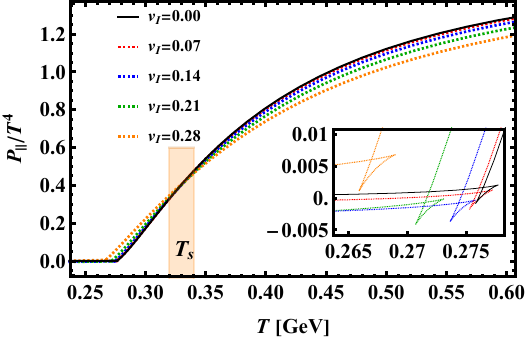}
    \includegraphics[width=.43\textwidth]{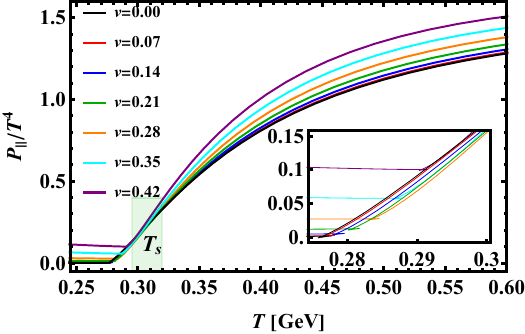}
    \caption{
    Temperature dependence of the longitudinal pressure $P_\parallel/T^4$ at different imaginary (\textbf{Left panel}) and real (\textbf{Right panel}) angular velocities. Imaginary rotation lowers the transition temperature, while real rotation raises it. At small real rotation, the pressure curves intersect above $T_c$, defining the supervortical temperature $T_s$ where the rotational response changes sign. For $v^2$ larger than $v_c^2\simeq0.1$, the intersections disappear and the pressure response becomes conventional at all temperatures.
    }
    \label{fig:pressure}
\end{figure}

\textit{Behavior of thermodynamic quantities under rotation}---
The sign change of moment of inertia is reflected in the longitudinal pressure (Fig.~\ref{fig:pressure}): the upper panel shows imaginary rotation lowering the transition, while the lower panel shows real rotation raising it. At small real rotation, the pressure curves intersect above $T_c$, defining $T_s$. For $v^2>v_c^2\simeq0.1$, the intersections disappear and the response becomes conventional.

\begin{figure}
\centering
\includegraphics[width=.43\textwidth]{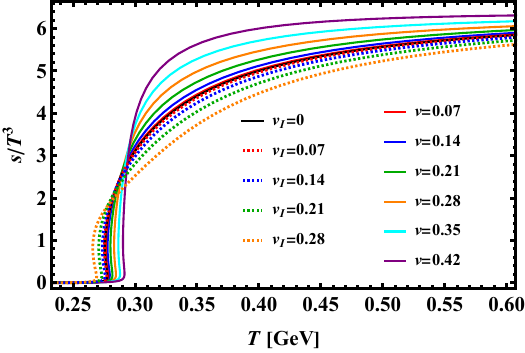}
\includegraphics[width=.43\textwidth]{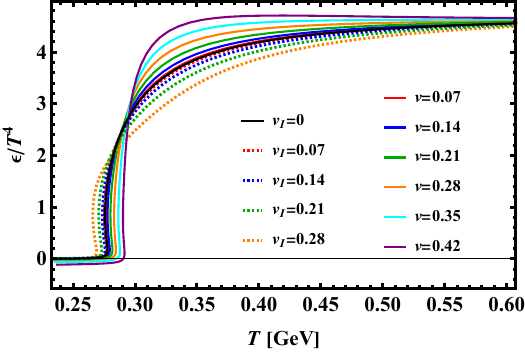}
\includegraphics[width=.43\textwidth]{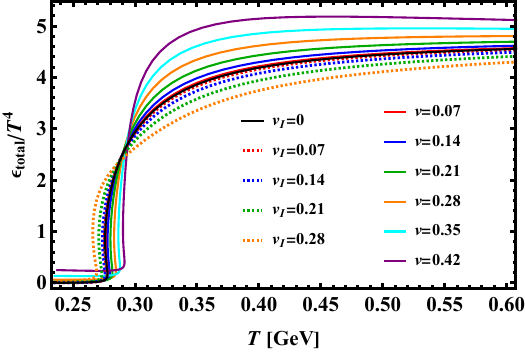}
\includegraphics[width=.43\textwidth]{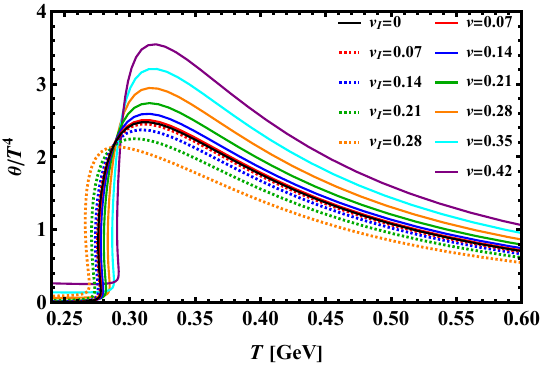}
\caption{Temperature dependence of thermodynamic quantities for pure gluonic matter at various angular velocities. The dashed and solid lines correspond to imaginary ($\Omega_I$) and real ($\Omega$) angular velocities, respectively. The four panels show the numerical results for Entropy density $s$, energy density $\epsilon$, total energy density $\epsilon_{\text{total}}$, and trace anomaly $\theta/T^4$, respectively. In the low-temperature phase, $\epsilon$ becomes negative at large real rotation speeds, but $\epsilon_{\text{total}}$, which includes the rotational term $\Omega J$, remains positive and increases with rotation. The shift of the critical temperature $T_c$ alters the rotational dependence of the thermodynamic quantities in the high-temperature phase.}\label{fig5}
\end{figure}
We calculate the entropy density $s/T^3$, energy density $\epsilon/T^4$, total energy density $\epsilon_{\text{total}}/T^4$, and the trace anomaly $\theta/T^4$ to study the thermodynamics of rotating pure gluonic matter. Fig.~\ref{fig5} shows these quantities as functions of temperature for both imaginary and real angular velocities. The behaviors of these thermodynamic variables are mainly affected by the shift of the critical temperature $T_c$ and the rotation-induced changes in the system.

In the low-temperature phase, $s/T^3$ remains nearly constant as rotation increases, while $\epsilon/T^4$ decreases. At large real angular velocities, $\epsilon/T^4$ becomes negative. However, the total energy density $\epsilon_{\text{total}}/T^4$, which includes the rotational contribution $\Omega J$, remains positive and increases with rotation across the calculated parameter space. In the high-temperature phase, the temperature dependence of $s/T^3$ and $\epsilon/T^4$ near $T_c$ is strongly influenced by the shift of $T_c$. Since imaginary rotation decreases $T_c$ and real rotation increases it, the values of $s/T^3$ and $\epsilon/T^4$ near the phase boundary increase with imaginary rotation but decrease with real rotation. At temperatures well above $T_c$, both quantities eventually decrease (increase) with imaginary (real) rotation in both cases.

The trace anomaly $\theta/T^4$ measures the deviation from the ideal Stefan-Boltzmann limit. In the low-temperature phase and at high temperatures ($T \gtrsim 0.3$ GeV), $\theta/T^4$ increases monotonically with rotation. Near the phase boundary, the peak structure of $\theta/T^4$ behaves differently for imaginary and real rotations. For imaginary rotation, the peak value decreases as rotation grows. For real rotation, the value of $\theta/T^4$ near $T_c$ decreases because $T_c$ shifts to higher temperatures, while the maximum peak height increases monotonically with rotation.

\end{document}